\documentclass[reprint, aip]{revtex4-1}

\usepackage{graphicx}
\usepackage{color}
\usepackage{multirow}
\usepackage{booktabs}
\usepackage{amsmath}
\usepackage{verbatim}
\usepackage{tikz}

\newcommand{\sub}[1]{\ensuremath{_{\rm #1}}} 
\newcommand{\super}[1]{\ensuremath{^{\rm #1}}} 

\newcommand{\nr}[0]{\ensuremath{n(\vec{r})}}

\begin{document}

\title{ Stability and surface diffusion at lithium-electrolyte interphases with connections to dendrite suppression }

\author{Yalcin Ozhabes}
\email[]{ao294@cornell.edu}
\altaffiliation{Contributed equally to this work}
\author{Deniz Gunceler}
\email[]{dg544@cornell.edu}
\altaffiliation{Contributed equally to this work}
\author{T. A. Arias}
\affiliation{Department of Physics, Cornell University, Ithaca, New York, 14853, USA}


\begin{abstract}
This work presents an \emph{ab initio} exploration of fundamental mechanisms 
with direct relevance to dendrite formation at lithium-electrolyte interfaces.  
Specifically, we explore surface diffusion barriers and solvated surface energies 
of typical solid-electrolyte interphase layers of lithium metal electrodes.
Our results indicate that surface diffusion is an important mechanism for understanding the recently observed
dendrite suppression from lithium-halide passivating layers, which were motivated by our previous work.
Our results uncover possible mechanisms underlying a new pathway for mitigating dendridic electrodeposition of lithium
on metal and thereby contribute to the ongoing efforts to develop stable lithium metal anodes for rechargeable battery systems.
\end{abstract}

\maketitle

\section{Introduction}

Development of more efficient energy storage technologies is needed to build a more sustainable future.
Understanding physical processes at the atomic scale on electrode-electrolyte interfaces 
is an important intermediate step towards realizing this goal.
Motivated by the desire to help enable many new applications, ranging from grid storage to long-ranged electric cars, 
researchers have also been using the tools of \emph{ab initio} electronic structure
to help develop better rechargeable lithium batteries.
\cite{materials-project,ictp,invited-paper,li-bare-surface,ceder,ceder2,intercalate,intercalate2,intercalate3,Li2CO3-diffusion,Li2CO3-diffusion-2,Li2CO3-diffusion-3,nishijima2014accelerated,ceder2011recharging,ong2011voltage,kang2006electrodes, KevinLeungMD, physrevapplied}

The current state-of-the-art in rechargeable batteries is lithium-ion technology, 
where the presence of a graphitic anode host results in deadweight (carbon) to be carried along with the battery.  
Metallic anodes would be a better choice due to their increased energy density ($\sim 3860$ mAh g\super{-1}), \cite{stanford-nature, samsung}
but they suffer from localized nucleation while charging and form needle-like structures called \emph{dendrites}.
\cite{samsung, goodenough-review, insitu-dendrite}
Despite many years of concentrated effort, 
there are still many unanswered questions about the
underlying physical mechanisms of lithium dendrite initiation and growth.  
This is partly due to the complex nature of the passivation layer, 
also called the solid-electrolyte interphase (SEI), 
that forms when the metallic electrode comes in contact with the electrolyte.
\cite{SEI-original, SEI-general, nmr-SEI}
Existing ideas and models on dendritic electrodeposition of lithium suggest that chemical inhomogeneities in the SEI layer result 
in spatially varying rates of deposition on the surface,
which then lead to instabilities because any protrusion tends to concentrate electric field lines.
\cite{Chazalvier, mukul, sei-macro, pulse-charge, afm-dendrite, macroReview, lithiumContaminantExp}

Recent experiments have shown that the composition of the SEI layer has a dramatic effect on the performance of the battery cell.
Based on our previous theoretical work, \cite{ictp, invited-paper}
Tu et. al. have managed to suppress dendrite growth in liquid and nanoporous electrolytes
by passivating the surface of the anode with lithium-halides. \cite{archer-nature}
Likewise, researchers from Stanford have succeded in designing an interfacial layer from carbon nanospheres 
which improves cycling efficiency. \cite{stanford-nature}
Understandably, these results and many others stimulate a strong interest in studying the fundamental physical mechanisms within the SEI layer,
and to date, many studies have investigated the bulk properties of these materials (e.g. bulk diffusion of Lithium).
\cite{Li2CO3-diffusion, Li2CO3-diffusion-2, Li2CO3-diffusion-3}

The rich physics at the interface between anode and liquid electrolyte
is much less understood, though there have been very promising new developments in this field as well. 
\cite{ictp, li-bare-surface, kent-abinitmd, stanford-nature}
Very recently, J{\"a}ckle and Gro{\ss} have published a comparative study of 
metallic Lithium, Sodium and Magnesium surfaces. \cite{li-bare-surface}
Their DFT calculations suggest that surface diffusion is significantly faster on magnesium metal than on lithium metal,
which may be important to understand why lithium forms dendrites while magnesium does not.
While this work is very important,
we believe (and the authors themselves also point out) that more investigation in this area is needed 
because that work does not address the presence and the effect of the electrolyte and, 
critically, the fact that metallic electrodes do not present pure surfaces to the electrolyte 
but rather complex non-metallic passivating layers known as SEI.

In this paper, we focus on the physical processes on the surfaces of various SEI materials for metallic lithium anodes. 
In particular, we provide an explanation of the physical mechanisms 
by which halogen additives (especially F\super{-}) to the electrolyte
suppress dendrites and improve cycling efficiency \cite{archer-nature,archer-new,archer-new-2,LiF-SEI,fec,fec2}.  
To this end, we utilize density functional theory to calculate surface cleavage energies 
and surface diffusion barriers for the most commonly reported SEI materials in the literature, 
\cite{sei-composition,sei-composition-2,sei-composition-3}
and then use these results to help understand the experimentally observed trends.
Our hope is that such understanding will accelerate the development of novel anode materials to improve battery performance.  
Our results support the growing belief \cite{ictp, invited-paper, mukul, archer-nature, li-bare-surface}
that anode materials with high surface energy and surface mobility are desirable.
We also detail our previous claim, \cite{ictp} recently supported by experiment, \cite{archer-nature}
that Lithium-halide SEI layers have these desirable properties and 
that they are effective in suppressing dendrite growth on metallic lithium anodes.

\section{Computational Methods}

To perform first principles DFT calculations, we use the open source JDFTx software\cite{JDFTx} 
which is based on the direct minimization of an analytically continued total energy functional. \cite{DFT-CG} 
Ultra-soft pseudopotentials \cite{uspp} from the GBRV library\cite{gbrv} are generated 
using the Vanderbilt pseudopotential code. \cite{vanderbiltgenerator}
To account for electronic exchange and correlation, we use the PBE flavour of generalized gradient approximation. \cite{pbe}
Throughout this methods section, we work in standard atomic units; i.e. bohr (B) for distances and hartree (H) for energies.
All results in the sections below will be presented in more familiar SI units.

For the Brillouin zone sampling of bulk units, we use a k-point grid of $4\times4\times4$ 
which we determined by converging the total energy to a level of 0.1 mH.
The energy cut-off for the plane wave basis was 20 Hartree which was also consistent with the same convergence threshold. a
The nuclei were relaxed until the root mean square of the forces were below 0.1 mH/Bohr. 

\begin{figure}
	\center{\includegraphics[width=1\columnwidth]{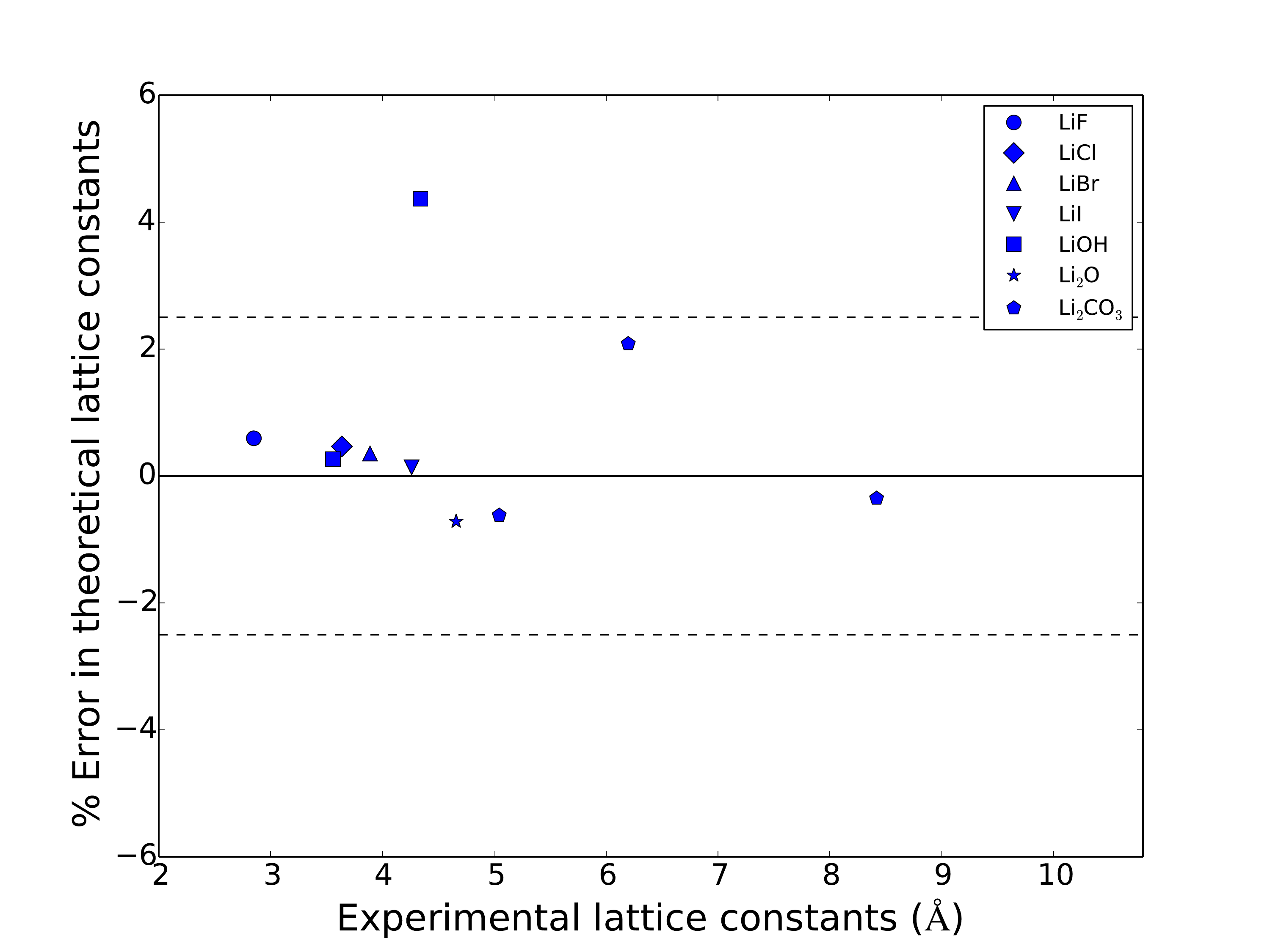}}
	\caption{
        The percentage error in DFT lattice constants, plotted as a function of the experimental lattice constants.
        The solid black line represents exact agreement of theory and experiment.
        The dashed black lines represent $\pm 2.5 \%$ deviations from experiment.
		\label{fig:bulk_properties}
	}	
\end{figure}

To test our choice pseudopotentials as well as other calculation parameters, 
we calculated lattice constants of various Lithium SEI materials.
The results, plotted in figure \ref{fig:bulk_properties}, were satisfactory.
(It is our belief that DFT has the largest error in the lattice constant of LiOH
because LiOH has a layered structure where long-ranged dispersion interactions play an important role.)

\begin{figure*}
	\center{
        \includegraphics[width=0.2\textwidth]{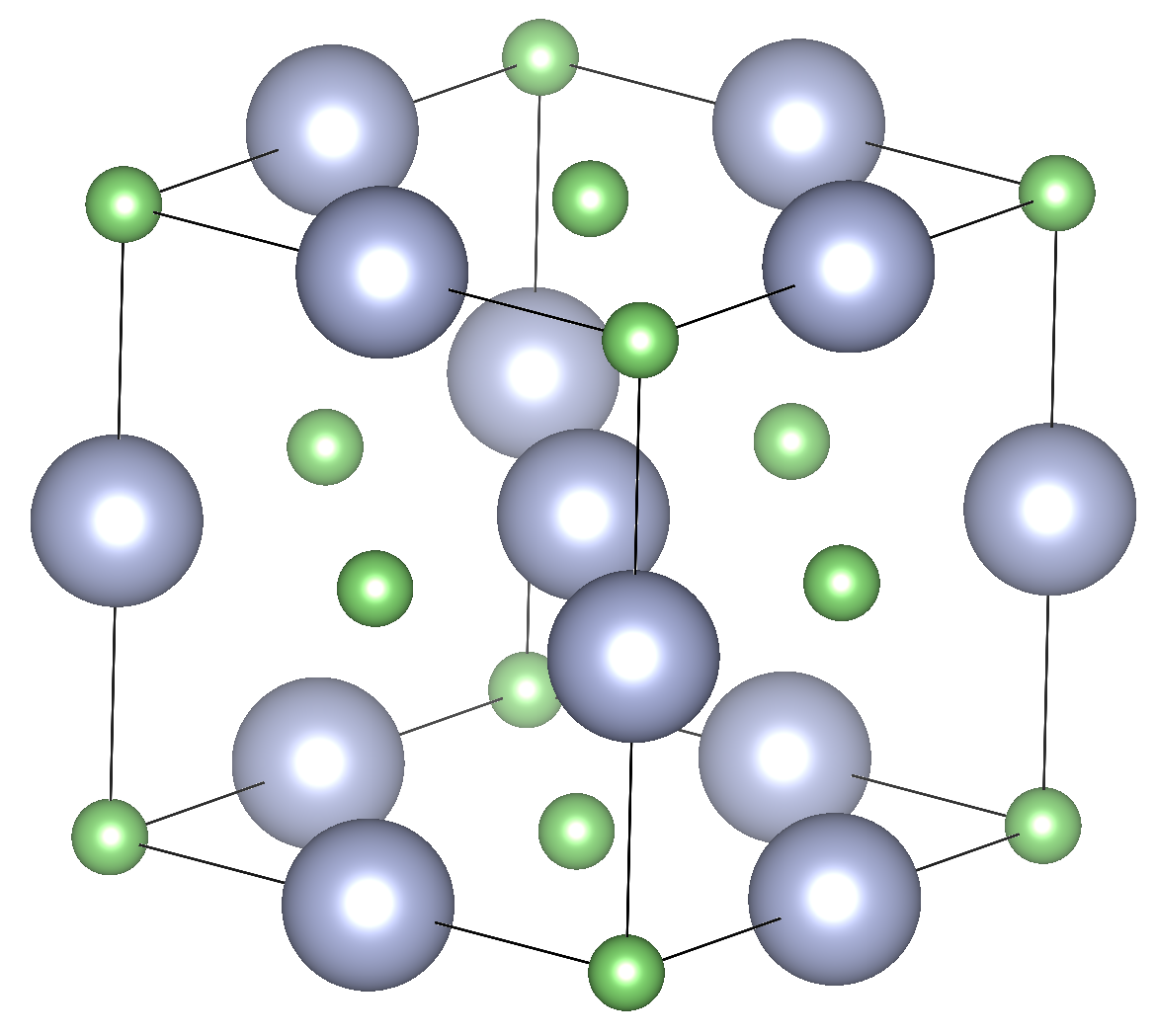} \hskip0.2cm
	    \includegraphics[width=0.2\textwidth]{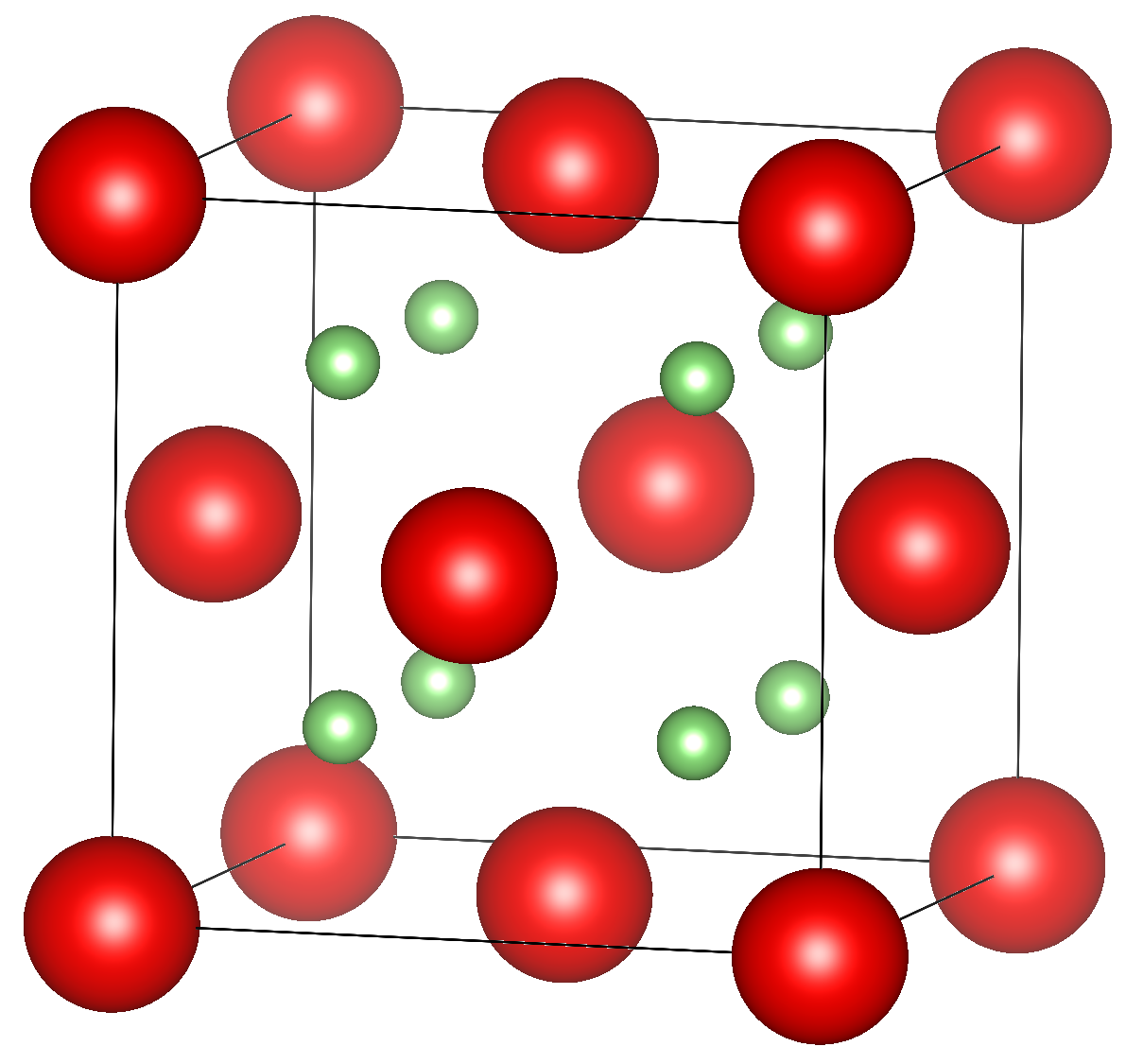} \hskip0.2cm
	    \includegraphics[width=0.25\textwidth]{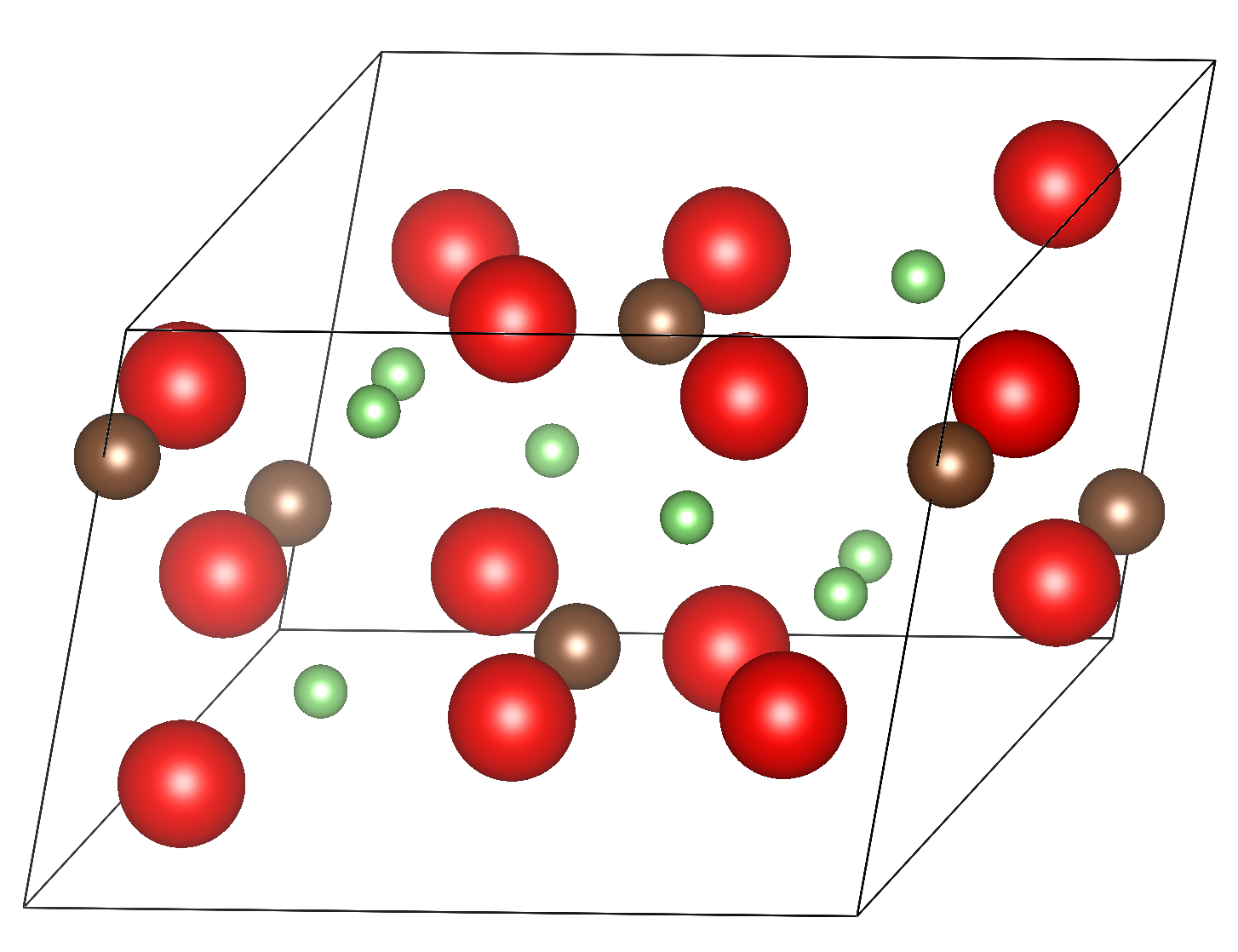} \hskip0.2cm
	    \includegraphics[width=0.2\textwidth]{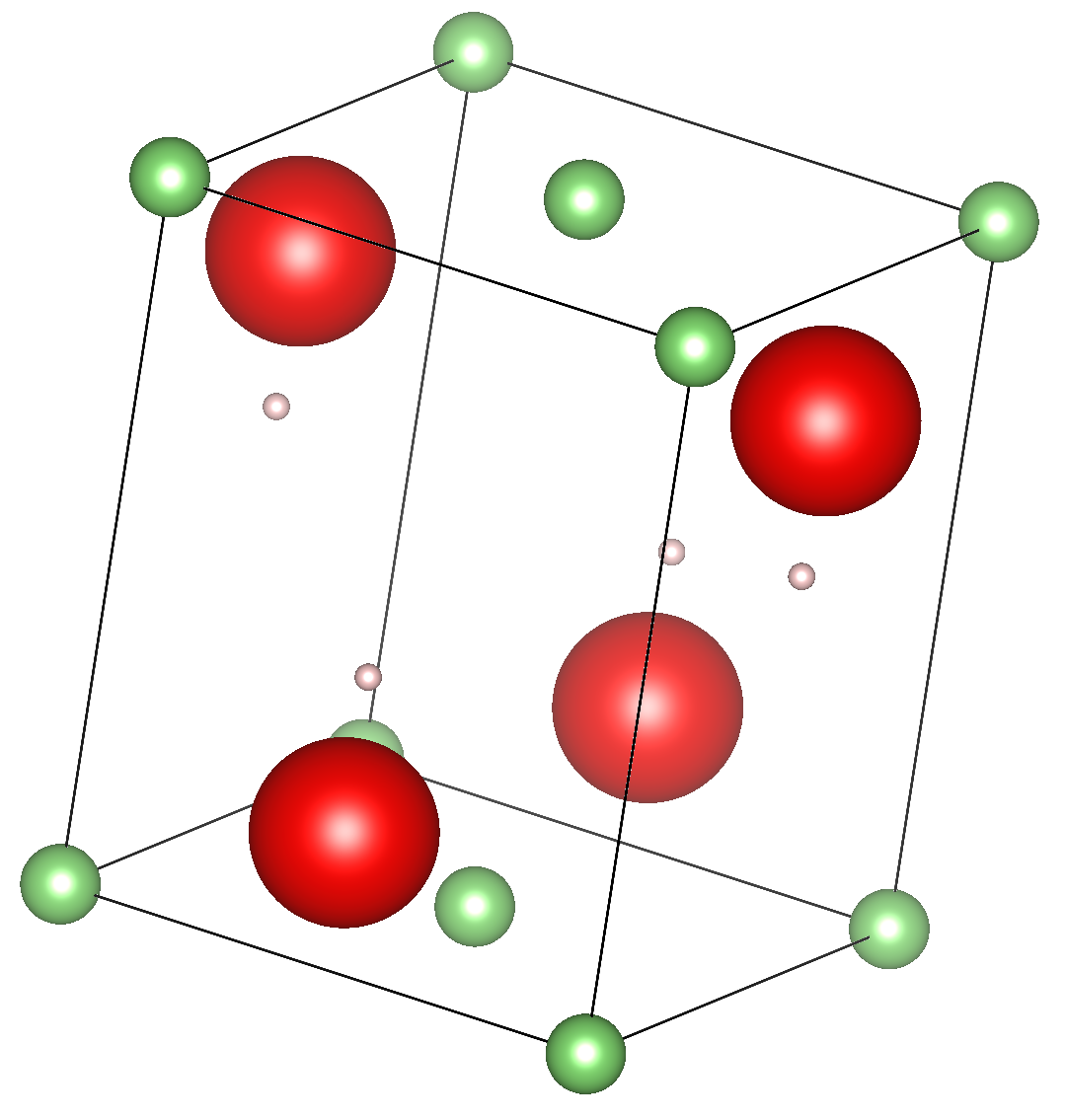}
    }
	\caption{
	    From left to right: rocksalt (Li-halides), anti-fluorite (Li\sub{2}O), zabuyelite (Li\sub{2}CO\sub{3}) and LiOH.
        Pictures were made using VESTA. \cite{vesta}
		\label{fig:structures}
	}	
\end{figure*}

To calculate the surface formation energies, we create slabs varying in size from compound to compound. To determine the thickness of each slab, we take the number of layers that are necessary to converge the surface energy to a level of 0.1 mH per unit cell. The vacuum layer (or solvent for fluid calculations) between the slabs is 20 Bohr, allowing us to collapse the three dimensional k-point grid to a planar grid. The center layer is held fixed and the rest of the slab is relaxed with the same convergence criteria mentioned for the bulk calculations. 
Where needed, we use a truncated Coulomb kernel \cite{coulomb-truncation} along the slab axis
to prevent spurious electrostatic interactions between slabs. 
Then we calculate surface formation energies 
by taking the difference of the slab energy and the energy of equal number of formula units in the bulk 
and then dividing by the surface area.
\begin{equation}
	E\sub{surf} = \frac{1}{2A}(E\sub{slab}-E\sub{bulk})
\end{equation}

To study surface diffusion, we consider the hopping process of an adatom that moves from one equilibrium position to the other and whose rate depends on the energy barrier along that path. We determine the diffusion path by comparing the binding energies of the adatom at the high symmetry points along the surface, letting the adatom relax in the direction parallel to the slab normal. Once we determine the endpoints of the diffusion pathway, we carry out a series of intermediate calculations by putting the adatom in a series of sites equally spaced on the line connecting the two binding sites. For these calculations, the entire slab relaxes except the middle layer, while the adatoms are restricted to stay on the plane perpendicular to the diffusion path.  To minimize the interactions between periodic images of adatoms, we use 3$\times$3 supercells.


Standard plane-wave electronic structure methods have difficulty handling interfaces between battery electrodes and the electrolyte, 
\cite{electrochemistry-is-hard}
largely due to the need to thermodynamically sample the configuration space of the liquid electrolyte.
As a result, there have been fewer \emph{ab inito} investigations of lithium metal anode-electrolyte interfaces compared to, for example,
the investigations of the bulk properties of lithium intercalation compounds.
\cite{ceder, ceder2, intercalate, intercalate2, intercalate3, physrevapplied}

In principle, one can sample the configuration space of the electrolyte using \emph{ab initio} molecular dynamics, \cite{kent-abinitmd, KevinLeungMD} 
and further accelerate the calculation using hybrid techniques such as QM/MM. \cite{qmmm}
However, calculation of free energies with these methods are difficult and 
these approaches often do not easily scale to the large number of materials we want to study.
An alternative approach is that of continuum solvation models, \cite{PCM-Marzari, invited-paper, general-glssa} 
where the individual molecules in the liquid electrolyte 
are replaced with a continuum field, and thus free energies can be computed with a single density-functional calculation.
Fortunately, recent developments in continuum solvation models
\cite{PCM-Marzari, invited-paper, anatase-pcm, general-glssa}
have made it feasible to efficiently study the solid-liquid interfaces for a large number of material/electrolyte combinations.

In this work, we use a nonlinear polarizable continuum model
which models the electrolyte environment as a continuous field of interacting dipoles. \cite{invited-paper}
This approach can also capture dielectric saturation effects, which are important near the surfaces of highly polar materials
such as the Lithium SEI surfaces we consider here.
In this nonlinear continuum model, the density profile of the electrolyte ($s(\vec{r})$) is computed self-consistently from 
the electron density of the surface slab ($n(\vec{r})$) as
\begin{equation}
    s(\vec{r}) = \textrm{erfc} \frac{\ln(\nr/n_c)}{\sigma\sqrt{2}},
\end{equation}
where $\sigma$ ($= 0.6$) determines the width of the transition region that is set to be resolvable on the FFT grid and 
$n_c$ is a (solvent-dependent\cite{general-glssa}) critical electron density value that determines the location of the solute-solvent interface.
For the non-electrostatic terms in the surface-electrolyte interaction, (such as cavitation entropy and long-ranged van der Waals)
we make use of the effective surface tension approximation, \cite{PCM-Marzari}
which modifies the bulk (macroscopic) surface tension of the electrolyte to approximate these contributions.

\section{Surface energies and diffusion}

The formation of dendrites represents an increase in surface area.
The thermodynamic perspective would thus indicate that SEI materials with greater surface energies should offer greater dendrite resistance.
Furthermore, dendrite nucleation may be driven by cracks in the SEI, \cite{samsung}
so that a stable SEI with a high surface formation energy would offer resistance to this mechanism as well.

However, the kinetics during electrodeposition might very well drive the system far from equilibrium.
Therefore, one must also consider the mechanism by which surface energy would tend to suppress dendrites, namely surface diffusion.
Following the same train of thought, one expects that materials with fast surface diffusion, 
for example those with small diffusion barriers for adatoms, would be less likely to form dendrites.

This section is broken into four parts:
In the first part, we obtain order-of-magnitude estimates 
for those diffusion barrier heights which would make diffusion an important process during electro-deposition.
We find that, indeed, most SEI materials in the following sections do have diffusion barriers in the relevant range.
In the second part, we investigate a binding-site switching mechanism that results in unusually low diffusion barriers for some lithium halides.
In the third part, we summarize our results for the surface energies and diffusion barrier heights for many candidate lithium SEI materials,
and discuss the trends we observe.
We investigate multiple classes of materials: 
rocksalt halides (LiF, LiCl, LiI, LiBr), layered (LiOH) and multivalent (Li\sub{2}O, Li\sub{2}CO\sub{3}).
In the final part of this section, 
we present an intriguing correlation between our diffusion barrier results and the experimentally observed time to short circuit.

\subsection{Comparing rates of deposition and diffusion}
cat: rates/main.tex: No such file or directory

It is important to know whether, at experimental conditions, surface diffusion can be competitive with electrodeposition.
We now probe this question with a simple order-of-magnitude analysis.
We assume that diffusion is a random walk where the hop rate is given by the Arrhenius equation $R = R_0 e^{- \frac{\Delta E}{kT} }$,
where $kT$ is thermal energy, $\Delta E$ is the diffusion barrier and $R_0$ is a base attempt rate related to the material's phonon frequencies.
To determine the balance between surface diffusion and deposition, let $q_e$ be the charge of the arriving ions, 
$J$ be the current density and $a$ be the dimension of the surface unit cell.
Under these conditions, we expect that dendrites form when the local current density becomes high enough to overcome surface diffusion,
specifically when the mean distance traveled by a diffusing atom ($d(t) = \sqrt{ R t } a$)
in the (mean) time interval between two deposition events at the same site
($t = q_e / \left( J a^2 \right)$) is smaller than the growing dendrite tip.
In this simple scenario, the critical barrier at which the rate of electrodeposition equals the rate of diffusion, is given by 
$\Delta E_c = 2 kT \ln{ \left[ \frac{1}{d} \sqrt{ \frac{R_0 q_e}{J} } \right] }$.

Experiments typically use current densities on the order of $1.0 \hskip5pt \rm{mA/cm^2}$,
but owing to the highly non-uniform rate of deposition on the surface, especially near any protrusion,
the relevant current densities can be orders of magnitude higher.
Indeed, measurements of lithium dendrite growth \cite{dendrite-tip,dendrite-tip-2}
have yielded the current densities at the dendrite tip in the $100-1000$ mA/cm\super{2} range.
As for $R\sub{0}$, following previous studies, \cite{li-bare-surface} we assume a standard phonon frequency of $\sim10$ THz.
For any microscopic protrusions in the order of 1-100 micrometers, 
we see that the critical diffusion barriers that result in competitive rates are of a few tenths of an electron volt.
At 100 mA/cm\super{2} current density the critical barrier for a protrusion of size 5 micrometers is approximately 0.1 eV,
which would mean that for such a system diffusion on a surface with a barrier above this value 
would be too slow to mitigate the growth of dendrites regardless of themodynamic unfavourability.
In fact, we do find that surface diffusion barriers on typical SEI materials 
are in this range and so a detailed study of surface diffusion is necessary for understanding dendritic processes.
Finally we note that, one must take this very simple analysis with care, as quantitative predictions other than order-of-magnitude estimates 
are likely to be beyond its capabilities.

\subsection{Change in the binding sites of Li-halides}

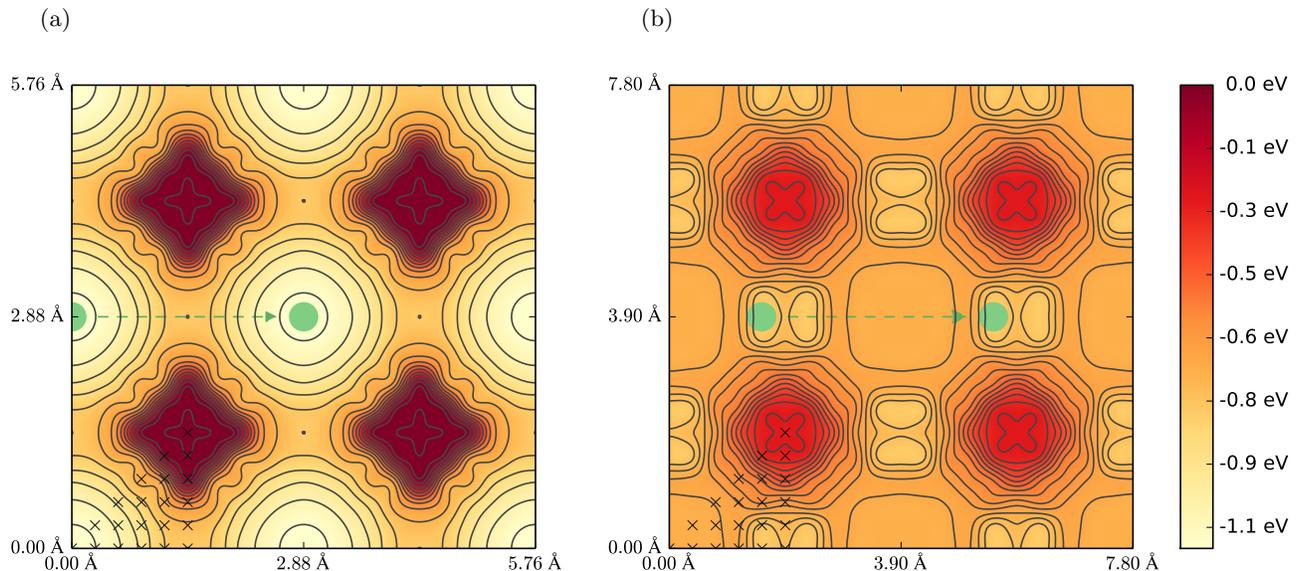
\begin{figure*}
\begin{tikzpicture}
\begin{scope}[xshift=-5cm]
	\node at (-8,0) {(a)};
	\node at (0,0) {(b)};
	\node at (8,0) { };
\end{scope}
\end{tikzpicture}
\center{\includegraphics[width=2\columnwidth]{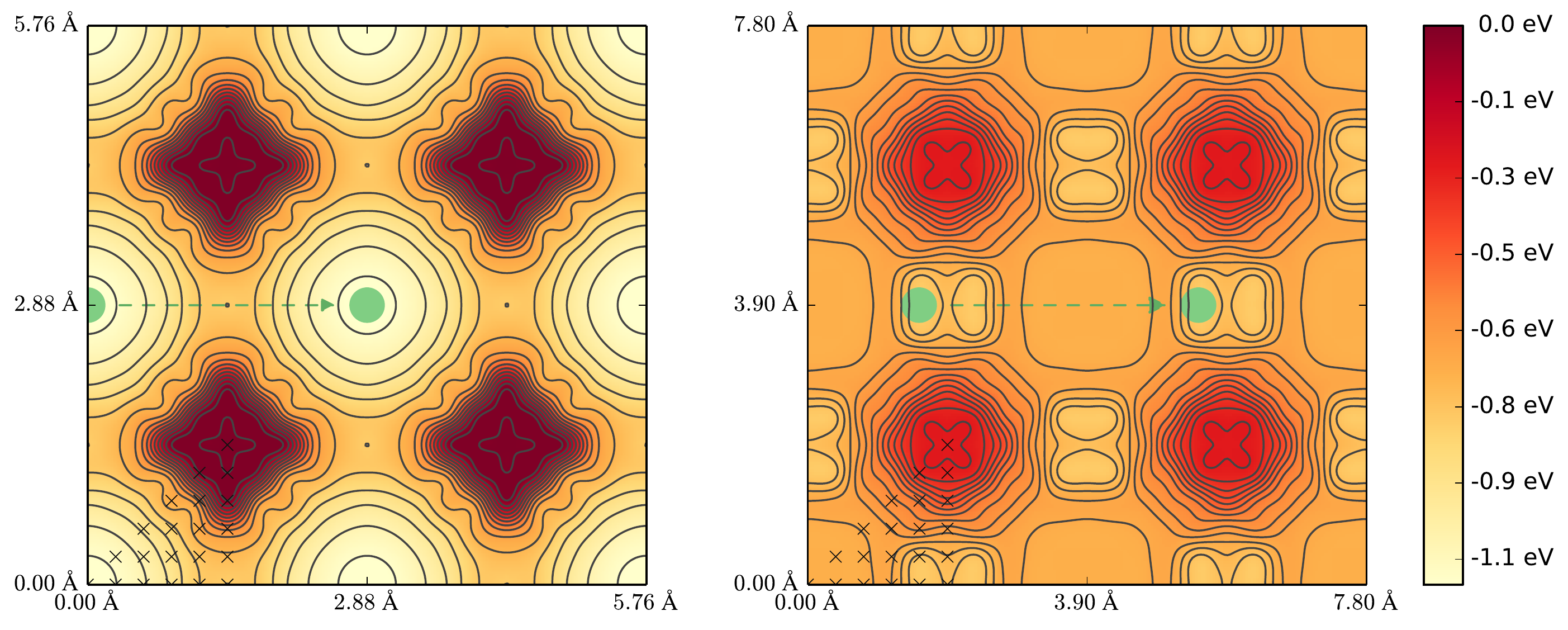}}
\caption{
Surface binding energies versus its binding site. 
LiF is in the left, LiBr is on the right. 
The lowest energy binding sites are indicated with a green circle and the diffusion path is as shown with the green arrows. 
The black cross markers indicate our data points, the whole contour plot is generated using symmetries of the surface and cubic interpolation.
\label{fig:halide-binding}
}
\end{figure*}

Below we find that of all the SEI materials we have studied, the lithium-halides are the most promising from the above point of view,
with diffusion barriers ranging from 0.03 to 0.15 eV.
The physical reason for these low barriers are as follows.
Even though the bulk structure of all the lithium-halide materials we consider here is the same (rocksalt), 
the binding site for adatoms changes as the anion size increases.
For halides with small anions (F and Cl) the binding site for the adatom
is directly above the anion (``anion site"), and in the transition state for diffusion,
the adatom sits between two anions and two cations (``in-between site").
See figure \ref{fig:halide-binding} (a) for an illustration.
On the contrary, for large anions, the roles of these two sites are reversed and the binding site sits at the ``in-between site"
(slightly off-center) while the transition state has the adatom at the ``anion site".
(See figure \ref{fig:halide-binding} (b))
Now, the diffusion barrier is equal to the energy difference between the binding site and the transition-site.
Because of this, halide surfaces that are in the neighbourhood of this change (LiCl and LiBr), where the sign of this difference changes, 
have very low diffusion barriers.
On the other hand, LiF, which is far from this change, have a relatively larger barrier, but still small compared to non-halide SEI materials.

The reason for the change in the binding site with anion size is steric interactions.
For the two smaller halogens, the in-between site is too close to the two cations,
which makes it energetically less favorable.
However, as the anions get bigger, the distance from the in-between site to the cations increases and
the adatom prefers to place itself in this halfway point where it can also interact with two anions simultaneously. 
Finally we note that even though the binding site changes as one goes from LiCl to LiBr, the diffusion path remains the same.

\subsection{Predictions for battery performance}

\begin{center}
\begin{figure}
\center{\includegraphics[width=1\columnwidth]{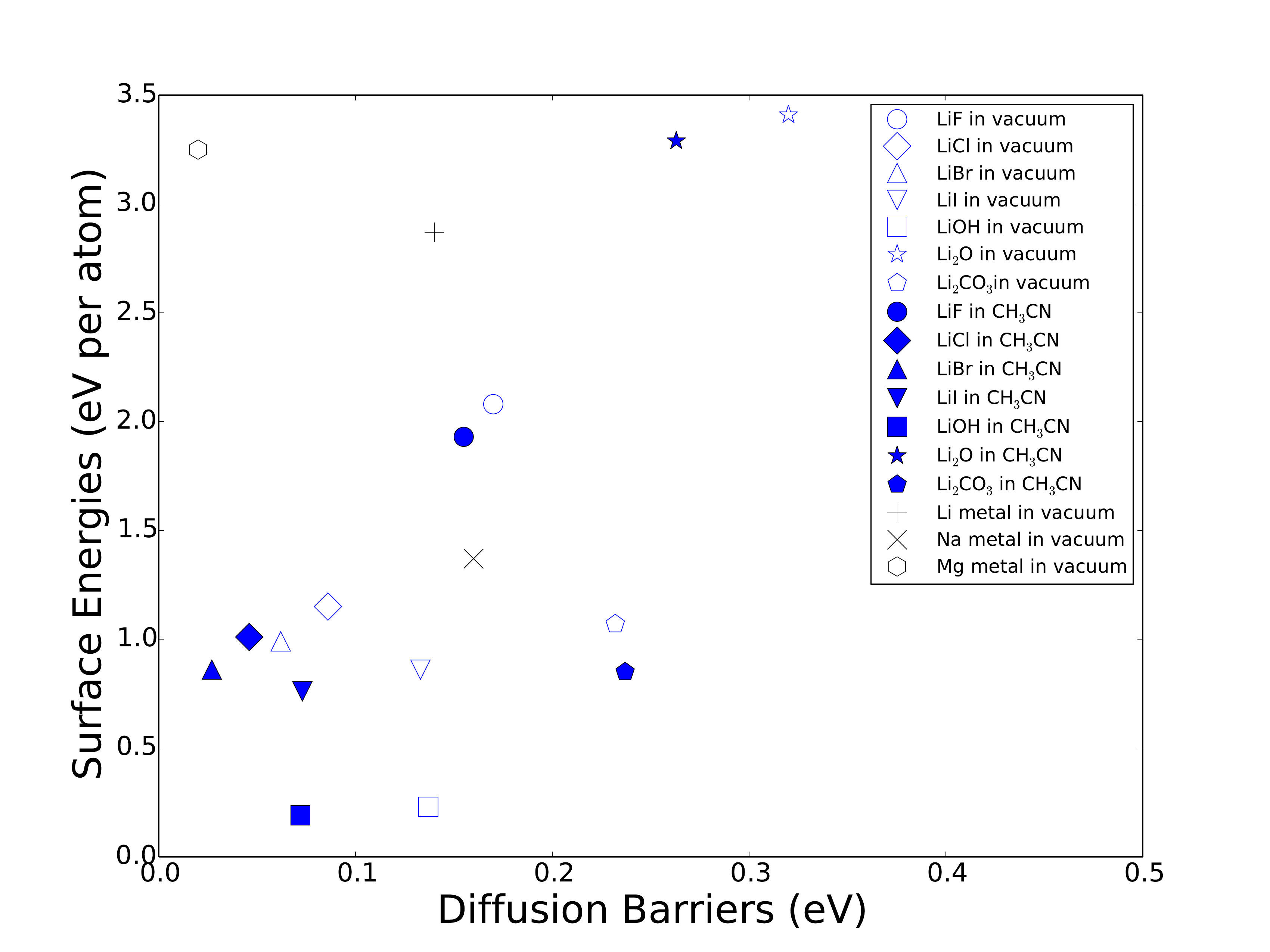}}
\caption{
    The surface diffusion barrier (x-axis) and the surface energy (y-axis) of various Lithium SEI materials.
    Hollow data points represent surfaces in contact with vacuum 
    whereas filled data points represent the same surfaces in contact with the electrolyte 
    (CH\sub{3}CN, modeled with a polarizable continuum model).
    The black markers indicate the bare metal calculations by J{\"a}ckle and Gro{\ss}. \cite{li-bare-surface}
    \label{fig:barrier_vs_energy}
}
\end{figure}
\end{center}

Figure \ref{fig:barrier_vs_energy} summarizes all of our results, 
not only for the diffusion barriers for all of our surfaces (including halides, as well as the hydroxide, carbonate and oxide) 
but also for the corresponding surface energies.
The data show that the presence of the electrolyte has significant impact, especially on the surface diffusion barriers.
The surface energies of all ionic crystals go down (often by $5-15\%$), 
owing to the strong electrostatic interaction between the surface and the solvent.
The surface diffusion barriers, however, change more dramatically, by up to as much as a factor of 2.
The diffusion barrier for all materials but one (Li\sub{2}CO\sub{3}) decrease when the electrolyte is included in the calculation.
These changes in diffusion barriers are significant because the rate depends exponentially on the barrier height
and because these energy changes are on the order of several $kT$ (which, at room temperature is approximately 0.025 eV).

The data also suggest a strong positive correlation between surface energies and surface diffusion barriers for most, 
but not all, SEI materials.
The most severe trend breakers are Magnesium metal (black hexagon), 
lithium metal (black plus) and LiOH (blue square).
Of these, the metals break the trend likely due to their very different electronic structure,
where Li is not in an oxidized (positively charged) state. 
LiOH likely breaks the trend because its layered structure and large intra-layer distance
cause it to have a very low surface energy along the z-axis.

Li\sub{2}CO\sub{3}, present in the SEI layer formed in the presence of many commonly used electrolytes 
(propylene carbonate, ethylene carbonate and others), is a less severe trend breaker.  
It has low surface energy and high diffusion barrier, both of which are undesirable quantities in a SEI material.
Lithium halides, on the other hand, have lower surface diffusion barriers than Li\sub{2}CO\sub{3} 
while also having either equal or higher surface energies.
Our hypothesis, first put forth in an earlier work, \cite{ictp} is that the low barriers may help explain
the experimentally observed phenomenon \cite{archer-nature} in which the formation of an lithium-halide SEI is effective in suppressing dendrites.
We further hypothesize that these mechanisms may be relevant 
not only in experiments where the electrolyte has been seeded with a Li-halide crystal, 
\cite{archer-nature} but also in experiments where other additives containing fluorine (e.g. hydrofloric acid or fluoroethylene carbonate) 
\cite{fec, fec2, LiF-SEI} have been used to improve stability and suppress dendridic growth.

As for LiOH and Li\sub{2}O, which are also occasionally observed in experiments, \cite{archer-nature}
LiOH appears to be undesirable because of its low surface energy, 
whereas Li\sub{2}O appears to be undesirable because of its high diffusion barrier.
However, the superiority of Li-halides over LiOH/Li\sub{2}O is not as conclusive 
because the halides are superior in only one of the two indicators.

Finally, among halides, figure \ref{fig:barrier_vs_energy} shows
that the stability (i.e. surface energy) decreases as one goes down the column of the peridic table, from F to Cl to Br to I. 
This is likely due to some combination of the decrease in the electronegativity of the ions (which weakens the strength of the ionic bonds)
and the steric interactions increasing the size of the lattice (which decreases the electrostatic stability of the lattice).
Decreased stability is an undesirable property in battery materials as it tends to lower the voltage at which the surface breaks down.

\subsection{Comparisons with measurements of short-circuit time}

\begin{figure}
\center{\includegraphics[width=1\columnwidth]{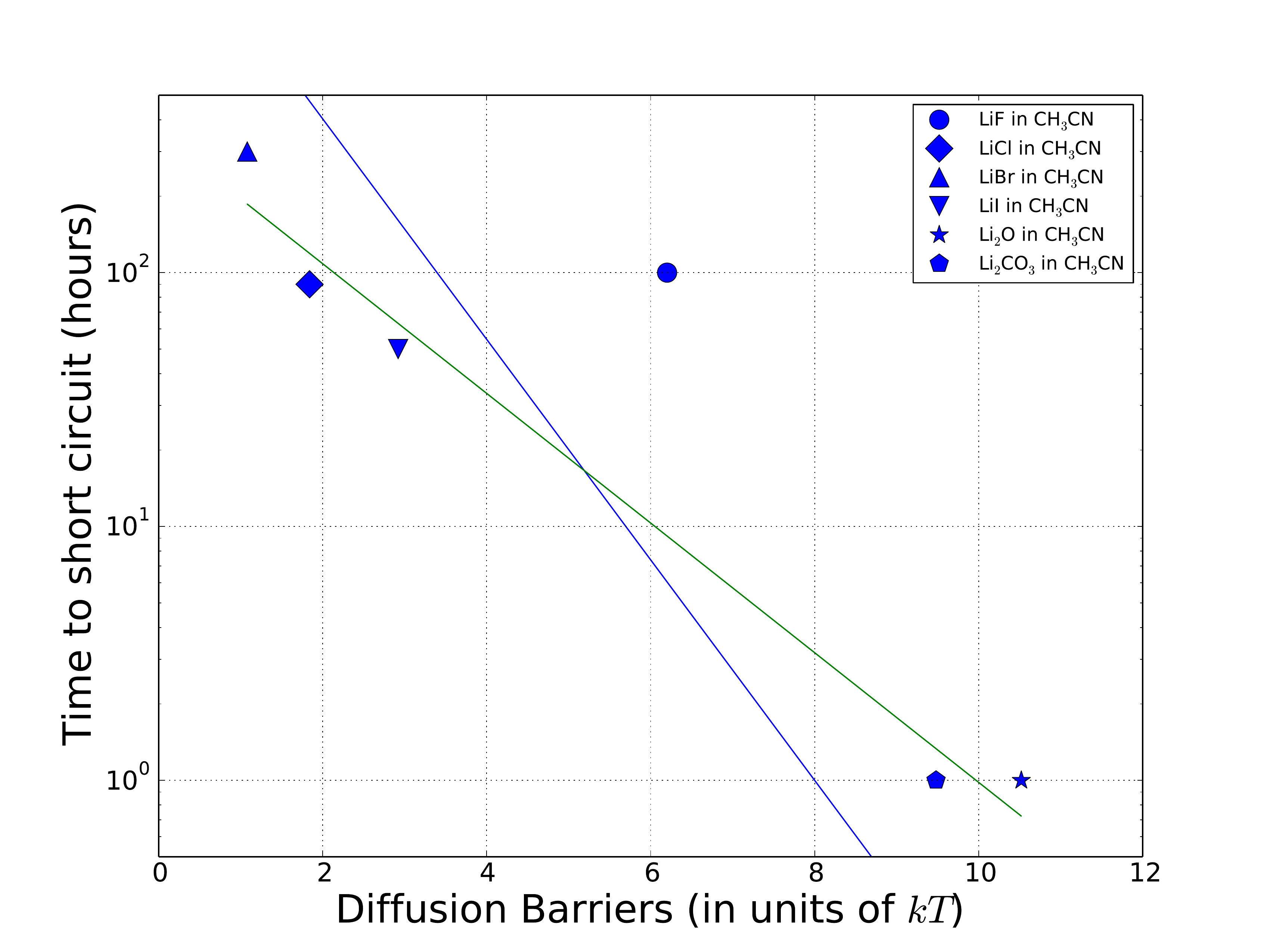}}
\caption{
Battery short circuit times (from \textcite{archer-nature}) plotted as a function of the surface diffusion barrier.
The green line is the best fit line, excluding the one outlier.
The blue line shows has the slope for an Arrhenius process.
}
\label{fig:Tsc_vs_F}
\end{figure}

From a practical perspective, the ultimate figure of merit is the length of time before the battery short circuits. 
Figure \ref{fig:Tsc_vs_F} plots this quantity, as measured by \textcite{archer-nature} in a symmetric cell,
as a function of our calculated diffusion barriers. 
Because diffusion is an activated process, we choose a logarithmic scale for the vertical axis.
Moreover, for Li\sub{2}CO\sub{3}, Li\sub{2}O and LiOH there are no separate data,
instead, for this single measurement that we have, where there are no halide additives,
the surface of the anode contains spatially-varying fractions of all three of these species.
The observed breakdown is likely due to the most-dendrite prone among these three materials, and thus represents
a lower bound for the lifetime of each of the materials.
Due to the notably lower barrier which we calculate for the hydroxide,
we strongly suspect that the observed cycle lifetime is associated either with the carbonate or the oxide.
Finally, because the barriers we calculate for these latter two materials are so close, 
we can not determine unambiguously which of these two materials fails first.  
Accordingly, the figure includes the experimental lifetime twice, once for each one of these two materials.

The data in figure \ref{fig:Tsc_vs_F} show a clear linear trend consistent with an Arrhenius-like behaviour.
However, the slope does not appear to be exactly unity, demonstrating the importance of other factors beyond the diffusion barrier.
Additionally, one point, lithium fluoride, appears to be an outlier, requiring additional explanation.
LiF has the highest barrier among the halides but the second longest time to short circuit.
We suspect that this deviation from the trend may be tied to it having the most stable surface among the halides
(see figure \ref{fig:barrier_vs_energy}), thus affording it additional stability.
Lithium fluoride's high surface energy derives not only from the high electronegativity of the fluoride atom,
but also from its very small lattice constant and surface area per formula unit.
(LiCl, the second smallest halide, has a surface area $\sim60 \%$ larger than LiF.)

While it is not suprising that a single quantity (the diffusion barrier) is not enough to fully explain the experimental lifetimes,
we find it encouraging that there is such strong correlation between diffusion barriers and the experimental lifetimes.


\section{Conclusion}

Recent experiments, prompted by earlier theoretical work, \cite{ictp} 
confirm the success of Lithium-halide additives in suppressing dendrite growth, \cite{archer-nature,archer-new,archer-new-2}
consistent with the findings of previous experiments with other fluoride-containing electrolyte additives. \cite{fec, fec2, LiF-SEI}
Prompted by this, we set out to explore more deeply the mechanisms of dendrite suppression at the atomic level.

We performed density-functional calculations to determine the surface energies 
and the surface diffusion barriers of lithium solid-electrolyte interphase materials.
Our calculations show that a lithium-halide SEI layer results in increased stability of the surface (higher surface energy) 
and faster diffusion along the surface (lower surface diffusion barrier for adatoms), 
both of which are likely important in explaining the above phenomenon.
Furthermore, our results provide an explanation for the unusually low diffusion barries on halide surfaces, 
tracing this effect back to a change in the binding site.  
This change in binding site, in turn, is driven by the trends in the electronegativity and the sizes of the anions in the halide crystals.
Finally, we observe a clear, approximately Arrhenius correlation between battery time to short circuit and surface diffusion barriers,
that can be used as a guide to suggest new material systems to further enhance stability against dendrite formation.

This work, which focused on solid-electrolyte interphase materials, leaves many directions yet to be explored.  
In mitigating dendrite growth, multiple diffusion pathways are available, including surface diffusion, 
bulk diffusion, and diffusion at the SEI-metal interface.  
A study similar to this one, but for the SEI-metal interface, would be illuminating.   
Finally, with the potential importance of surface diffusion established, 
results from \emph{ab initio} studies should be incorporated into macroscopic models of dendrite growth.

\begin{acknowledgements}
The authors would like to thank Prof Lynden Archer, Prof Ersen Mete, Mukul Tikekar, Pooja Nath 
and Snehashis Choudhury for encouragement and useful discussions,
and Christine Umbright for her help in writing this manuscript.
This work was supported as a part of the Energy Materials Center at Cornell (EMC\super{2}), an Energy Frontier Research Center funded by the U.S. Department of Energy, Office of Science, Office of Basic Energy Sciences under Award Number DE-SC0001086.
\end{acknowledgements}

\bibliography{references.bib}
\end{document}